\documentclass[11pt]{article}
\usepackage{latexsym}

\usepackage{amssymb}
\usepackage{amsmath}

\numberwithin{equation}{section}

\newcommand{\be}{\begin{equation}}
\newcommand{\ee}{\end{equation}}
\newcommand{\ba}{\begin{eqnarray}}
\newcommand{\ea}{\end{eqnarray}}
\newcommand{\bi}{\begin{itemize}}
\newcommand{\ei}{\end{itemize}}
\newcommand{\baa}{\begin{array}}
\newcommand{\eaa}{\end{array}}

\newcommand{\nr}[1]{(\ref{#1})}

\newcommand{\rmi}[1]{{\mbox{\scriptsize #1}}}

\newcommand{\fra}[2]{{\textstyle{\frac{#1}{#2}\,}}}  
\newcommand{\mn}{{\mu\nu}}

\newcommand{\hr}{{\hat r}}
\newcommand{\hrp}{{\hat r_+}}

\newcommand{\hz}{{\hat z}}

\newcommand{\heps}{{\hat\epsilon}}

\def\Tr{{\rm Tr\,}}

\def\CL{{\cal L}}

\def\CN{{\cal N}}
\def\gsim{\raise0.3ex\hbox{$>$\kern-0.75em\raise-1.1ex\hbox{$\sim$}}}
\def\lsim{\raise0.3ex\hbox{$<$\kern-0.75em\raise-1.1ex\hbox{$\sim$}}}

\DeclareMathOperator{\sgn}{sgn}

\newcommand{\edoc}{\end{document}}

\begin{document}

\begin{titlepage}
\begin{flushright}
HIP-2008-28/TH\\
\end{flushright}
\begin{centering}
\vfill

{\Large {\bf Isotropic AdS/CFT fireball}}

\vspace{0.8cm}

\renewcommand{\thefootnote}{\fnsymbol{footnote}}

K. Kajantie$^{\rm a,b}$\footnote{keijo.kajantie@helsinki.fi},
Jorma Louko$^{\rm c}$\footnote{jorma.louko@nottingham.ac.uk},
T. Tahkokallio$^{\rm a,b,d}$\footnote{ttahko@uvic.ca}

\setcounter{footnote}{0}

\vspace{0.8cm}

{\em $^{\rm a}$%
Helsinki Institute of Physics, P.O.Box 64, FI-00014 University of
Helsinki, Finland\\}
{\em $^{\rm b}$%
Department of Physics, P.O.Box 64, FI-00014 University of Helsinki,
Finland\\}
{\em $^{\rm c}$%
School of Mathematical Sciences, University of Nottingham,
Nottingham NG7 2RD, UK\\}
{\em $^{\rm d}$%
Department of Physics and Astronomy, University of Victoria, 
Victoria, British Columbia, V8W 3P6 Canada\\}

\vspace*{0.8cm}

\end{centering}

\noindent
We study the AdS/CFT thermodynamics of the spatially
isotropic counterpart of the Bjorken similarity flow in
$d$-dimensional Minkowski space with $d\ge3$, and of its
generalisation to linearly expanding $d$-dimensional
Friedmann-Robertson-Walker cosmologies with arbitrary values
of the spatial curvature parameter~$k$.
The bulk solution is a nonstatic foliation
of the generalised Schwarzschild-AdS 
black hole with a horizon of constant curvature~$k$. 
The boundary matter is an expanding perfect fluid that satisfies 
the first law of thermodynamics for all values of 
the temperature and the spatial curvature, 
but it admits a description as a scale-invariant
fluid in local thermal equilibrium only when the 
inverse Hawking temperature is negligible compared 
with the spatial curvature length scale. 
A~Casimir-type term in the holographic energy-momentum tensor is 
identified from the threshold of 
black hole formation and is shown to 
take different forms for $k\ge0$ and $k<0$. 

\vfill \noindent

\vspace*{1cm}

\noindent
Revised October 2008

\vspace*{1ex}

\noindent
Published in 
Phys.\ Rev.\ D {\bf 78}, 126011 (2008)

\vfill

\end{titlepage}

\section{Introduction}

AdS/CFT duality permits one to obtain the thermodynamics of a
conformal field theory (CFT) from classical solutions to
higher-dimensional gravitational field equations, provided the CFT
spacetime is attached to the gravitational solution as a conformal
boundary under suitable asymptotic conditions. The prototype
prediction is the equation of state of static $\CN=4$ Super-Yang-Mills
matter in Minkowski space~\cite{gkp,pss}, obtained from the
thermodynamics of an appropriate higher-dimensional black hole
solution. Currently there is a growing interest in situations where
the CFT configuration is not static and the duality yields predictions
for hydrodynamical transport
coefficients~\cite{nastase,ssz,Son:2007vk}. A~case of particular
interest is CFT as an approximation to QCD plasma in $(3+1)$ spacetime
dimensions, in a kinematical configuration that approximates a head-on
ion collision at relativistic velocities
\cite{jp,jp2,nakamura1,nakamura2,janik1,janik2,kovchegov,Lin:2006rf,%
Benincasa:2007tp,Alsup:2007bs,%
Heller:2008mb,Kinoshita:2008dq} 
or otherwise invokes space-time 
dependence \cite{Bhattacharyya:2007jc,Natsuume:2007ty,Natsuume:2007tz}. 

In the $(3+1)$-dimensional ion collision setting the kinematics
distinguishes between the longitudinal spatial dimension, in which the
original ion velocities are, and the two transverse spatial
dimensions. This spatial anisotropy is an essential part of the
physics of the situation, but it is also a source of significant
mathematical subtlety, and the bulk geometry is at present known only
in terms of asymptotic
expansions~\cite{Heller:2008mb,Kinoshita:2008dq}. By contrast, in the
simplified setting of $(1+1)$-dimensional ion collisions, where no
spatial anisotropy can occur, the bulk geometry can be found
explicitly~\cite{skendesolo,Kajantie:2008rx},
and in the exactly boost-invariant
special case the boundary fluid turns out to an 
expanding ideal gas in local thermal equilibrium~\cite{klt}.

In this paper we consider a setting in which the CFT lives on a
$d$-dimensional spacetime with $d\ge3$ and the fluid is expanding
isotropically from a point-like explosion. The case of main
interest is $d$-dimensional Minkowski spacetime, but it will turn out
useful to consider at the same time the whole family of linearly
expanding Friedmann-Robertson-Walker (FRW) cosmologies with arbitrary
real values of the spatial curvature parameter~$k$,
with the isotropically
expanding fluid in Minkowski space obtained as the special case
$k=-1$. The bulk solution can be found explicitly and is
the generalised Schwarzschild-AdS${}_{d+1}$
black hole~\cite{gibbons},
\be
ds^2 = -F dt^2 + \frac{dr^2}{F} + r^2 d\Omega_{d-1}^2(k),
\qquad
F   =
{(r/\CL)}^2+k- \frac{\mu}{{(r/\CL)}^{d-2}} ,
\label{eq:intro-bulk}
\ee
where $\CL$ is the length scale set by the bulk cosmological
constant and the rest of the notation is explained in
Section~\ref{sec:ads-schw-arb} and in the Appendix.
We find:

\begin{itemize}
\item
We first address (Sections \ref{sec:static}
and \ref{sec:static-thermo})  
AdS/CFT thermodynamics on the conformal boundary of
the bulk solution in the static foliation~\nr{eq:intro-bulk},
in which the conformal boundary is a static FRW
cosmology.
While some of
this material is well known
\cite{bala-kraus,Horowitz:1998ha,Emparan:1999pm,Emparan:1999gf,Burgess:1999vb}, 
the novel feature is that keeping $k$ general allows the
boundary volume $V$ to be treated as a continuous thermodynamical
variable, since $V\sim {\bigl(\CL/\sqrt{|k|}\bigr)}^{d-1}$ for $k\ne0$.
We find that the boundary fluid satisfies the first law of
thermodynamics in the usual form,
\be
dE=T\,dS-p\,dV .
\label{eq:firstlaw-intro}
\ee
However, the relation
\be
E=TS-pV,
\label{eq:scaling-relation}
\ee
which would usually be expected to hold for a
fluid in local thermal equilibrium, is satisfied only in the special
case $k=0$, in which the FRW cosmology is flat.
For $k\ne0$, \nr{eq:scaling-relation}
is satisfied
asymptotically when the bulk black hole is so large
that the thermal length scale $1/T$ is negligible compared with the
curvature length scale $\CL/\sqrt{|k|}$.
The reason for \nr{eq:scaling-relation} not being satisfied
for $k\ne0$ is that
the presence of spatial curvature breaks scale-invariance in the
fundamental thermodynamic relation of the fluid.
\item
We then write (Sections \ref{sec:tau-in-bulk} and \ref{sec:exp-thermo}) 
the bulk solution \nr{eq:intro-bulk} in a foliation that
makes the boundary metric into a
FRW cosmology with
a linearly expanding scale factor and analyse the AdS/CFT
thermodynamics of the boundary fluid.\footnote{For $d=4$, 
a generalisation of this coordinate transformation 
to an arbitrarily-expanding FRW cosmology is given in~\cite{siopsis}. 
Related discussions from the brane world perspective are given in 
\cite{Kraus:1999it,Kehagias:1999vr,Kiritsis:2005bm}.}
For $k=-1$ the fluid expands
isotropically and boost-invariantly from a point-like explosion in
a spacetime that is flat, but just written 
in a cosmological form that is adapted to the explosion; 
for $k\ne-1$, by contrast, the spacetime is a 
genuinely curved expanding cosmology, and the explosion occurs at a pointlike 
initial curvature singularity.
The first law is satisfied in the usual
form~\nr{eq:firstlaw-intro}, both when the differentials refer to the
time-evolution due to the expansion, and also when the differentials
refer to varying the spatial volume and the temperature at a fixed
cosmological time by varying $k$ and $\mu$ in the bulk
solution. However, the relation \nr{eq:scaling-relation}
is again satisfied exactly only for $k=0$, and for $k\ne0$ it is satisfied
asymptotically in the limit of large bulk black hole. The
reasons are the same as in the static case.
\end{itemize}

The fact that \nr{eq:scaling-relation} is not satisfied
for $k\ne0$ should not be surprising from the
bulk viewpoint. The
bulk counterpart of the boundary spatial curvature is the
curvature of the black hole horizon, and it
has been observed both within \cite{ss} 
and without \cite{Gibbons-Hawking} the AdS/CFT context that
horizon curvature can interfere with the description of matter as a
fluid in local thermal equilibrium. A~perhaps surprising feature in
the nonstatic boundary case is, however, that the relation
\nr{eq:scaling-relation} is not satisfied for $k=-1$, when the fluid
is just expanding
isotropically in Minkowski spacetime. In this case the boundary
curvature that breaks the scale invariance is not that of the spacetime in
which the fluid expands, but just the intrinsic curvature of the
boost-invariant spacelike hyperboloids that are surfaces of constant
proper time in the frame comoving with the fluid.

Owing to the spatial isotropy that we assumed at the outset, the fluid
flow has a vanishing shear tensor both in the static and non-static
cases. The shear viscosity does therefore not contribute to the
energy-momentum tensor, and we cannot use the system to examine the
usual AdS/CFT prediction $\eta=s/(4\pi)$ for the shear viscosity
$\eta$ and the entropy density~$s$~\cite{mateos-lectures}. In the
non-static case the fluid flow does have a nonvanishing expansion, and
the kinematics would hence allow a bulk viscosity term to appear in the
energy-momentum tensor, but the coefficient of this term vanishes
simply because our bulk action is pure Einstein and does not
contain fields that could break conformal invariance on the boundary.

A conceptual issue with the boundary matter is whether
the holographic energy-momen\-tum tensor $T_\mn$ 
should be attributed 
to the boundary fluid in its entirety, 
or whether it should be split as 
$T_\mn =
T_{\mu\nu}^{(\mathrm{Casimir})} + T_{\mu\nu}^{(\mathrm{fluid})}$,
where $T_{\mu\nu}^{(\mathrm{Casimir})}$ is a 
Casimir-type vacuum energy-momentum tensor. 
For example, 
a conformal scalar field in four-dimensional flat spacetime 
has a nonvanishing energy-momentum tensor
in the vacuum state adapted to 
the isotropic expansion~\cite{bunch-hyperbolic}, 
and this tensor turns out to share 
the tensorial structure and 
time-dependence of our holographic 
energy-momentum tensor. Similarly, 
a conformal 
scalar field in two-dimensional flat spacetime 
in the analogous vacuum state has a 
nonvanishing energy-momentum tensor
(\cite{bd}, equation (7.24)), 
and this tensor shares the 
tensorial structure and time-dependence of the 
holographic energy-momentum tensor for the 
$(1+1)$-dimensional Bjorken flow~\cite{klt}. 
When 
$T_{\mu\nu}^{(\mathrm{Casimir})}$ is chosen so that 
$T_{\mu\nu}^{(\mathrm{fluid})}$ vanishes 
at the threshold of a disappearing 
bulk black hole, 
the fluid in the $(1+1)$-dimensional Bjorken 
flow was found to be scale invariant~\cite{klt}, 
and we shall 
adopt this normalisation also in the present paper. 
We shall see, however, that the core conclusions of consistency of 
\nr{eq:firstlaw-intro}
for all $k$ and the 
inconsistency of 
\nr{eq:scaling-relation} for $k\ne0$ are 
not sensitive to the choice of $T_{\mu\nu}^{(\mathrm{Casimir})}$ 
and hold even if one chooses for example 
$T_{\mu\nu}^{(\mathrm{Casimir})}=0$. 
Prospects of identifying a Casimir term in the case of the 
usual, non-isotropic Bjorken flow are discussed in~\cite{klt4}. 

In the limit $d\to2$, our formulas reproduce the $d=2$ results
of~\cite{klt}.
The qualitative difference that appears in this limit is
that for the fluid expanding in Minkowski space, the relation
\nr{eq:scaling-relation} holds for $d=2$ but not for $d\ge3$. From
the viewpoint of this limit,
the exact solvability of the case $d=2$ can be
attributed to the absence of spatial anisotropy in two dimensions,
rather than just to the local triviality of $(2+1)$-dimensional Einstein
gravity \cite{henneaux,carlip}.

The metric signature we use is
mostly plus,
$({-}{+}{+}\cdots{+})$, and we use the $({+}{+}{+})$ sign convention
of~\cite{MTW}, except that in the single formula
\nr{regaction} the conventions are those of~\cite{skenderis}.
The boundary dimension is $d$ and the bulk dimension is
$d+1$. For simplicity of the presentation we assume throughout
$d>2$,
although most of the formulas remain valid also for $d=2$ when
interpreted in a proper limiting sense,
reducing to those given in~\cite{klt}.
The Lorentz-signature 
Einstein-Hilbert-York-Gibbons-Hawking action 
is \cite{Gibbons-Hawking,York:1972sj}
\be
S= \frac{1}{16\pi G_{d+1}} \int_M d^{d+1}x \, \sqrt{-g}
\left(R+ \frac{d(d-1)}{\CL^2}\right)
\ \ - 
\frac{1}{8\pi G_{d+1}}
\int_{\partial M} 
d^{d}x \, \sqrt{-\gamma} \, K(\gamma)  ,
\label{einstein-action}
\ee
and the bulk field equations are 
\be
R_{MN}=-{d\over\CL^2}g_{MN},
\label{adseqs}
\ee
where the positive constant $\CL$ is the length 
scale of the cosmological constant. 
To compute the holographic energy-momentum tensor~\cite{skenderis},
the bulk metric will be written in the Fefferman-Graham form
\be
ds^2=g_{MN}dx^M dx^N={\CL^2\over z^2} [ g_\mn dx^\mu dx^\nu+dz^2 ],
\label{fefferman}
\ee
where the boundary is at $z\to0$ and
$g_\mn$ has the small $z$ expansion
\be
g_\mn =g^{(0)}_\mn
+ g^{(1)}_\mn z
+ \cdots
+g^{(d)}_\mn z^d
+ \textrm{(logarithmic term)}
+ \cdots .
\label{gexp}
\ee
For odd $d$, the holographic energy-momentum tensor
on the conformal boundary metric $g^{(0)}_\mn$ is
\be
T_{\mu\nu} =
\frac{d \CL^{d-1}}{16\pi G_{d+1}} g^{(d)}_\mn .
\label{Tmunu-odd}
\ee
For even $d$ there are in general additional
terms~\cite{skenderis}. Those for $d=4$ are given below
in~\nr{eq:Tmunu-d=4}.

\section{AdS$_{d+1}$ black hole with constant curvature horizon}
\label{sec:ads-schw-arb}

In this section we review the thermodynamics of the
generalised Schwarzschild-AdS$_{d+1}$ black hole in the case of an
arbitrary constant curvature horizon.

The starting point is the bulk solution \cite{gibbons}
\begin{subequations}
\label{adsbh}
\begin{align}
ds^2 &= -F dt^2 + \frac{dr^2}{F} + r^2 d\Omega_{d-1}^2(k),
\label{adsbh-justmetric}
\\
F (\hr) & =
\hr^2+k-{\mu\over\hr^{d-2}},
\label{eq:Ffunc-def}
\end{align}
\end{subequations}
where $\hr = {r/\CL}$, $\mu$ is a dimensionless parameter and
$d\Omega_{d-1}^2(k)$ is a $(d-1)$-dimensional positive definite metric
of constant curvature, with the Ricci scalar normalised
to $(d-1)(d-2)k$ \cite{wolf}. As $d\Omega_{d-1}^2(k) = {|k|}^{-1}
d\Omega_{d-1}^2(k/|k|)$ for $k\ne0$, it would be possible 
(and is indeed conventional) 
to normalise
any nonzero $k$ to $k/|k| = \pm1$
by a redefinition of $t$ and~$r$,
but we shall keep $k\in\mathbb{R}$ in order to treat the spatial
volume as a thermodynamical
boundary variable in Section~\ref{sec:static-thermo}.
Relevant properties of $d\Omega_{d-1}^2(k)$ are reviewed in the Appendix.

The topology of the space~$M_{d-1}$, on which $d\Omega_{d-1}^2(k)$ is
the metric, will not play an essential role in what follows, and the
simplest choice is to take $d\Omega_{d-1}^2(1)$ to be the unit
sphere, $d\Omega_{d-1}^2(-1)$ to be the unit hyperbolic space
and $d\Omega_{d-1}^2(0)$
to be Euclidean~$\mathbb{R}^{d-1}$, as done in the Appendix.
We will however be using
thermodynamical quantities that
are defined with respect to unit volume of
$d\Omega_{d-1}^2(k)$, and it will reduce the verbiage to regard
$d\Omega_{d-1}^2(k)$ as a metric on a compact space for all values
of~$k$, with the finite volume denoted by~$\Omega_{d-1}(k)$. For
$k\ne0$, the scaling of the metric with $k$ implies $\Omega_{d-1}(k) =
{|k|}^{-(d-1)/2} \Omega_{d-1}(k/|k|)$, while $\Omega_{d-1}(0)$ may
take arbitrary values~\cite{wolf}.

We denote the coordinates on $M_{d-1}$ by
$y^i$, $i=1,\ldots,d-1$, and write
$d\Omega_{d-1}^2(k) = h_{ij} dy^i dy^j$.

The metric \nr{adsbh} is asymptotically locally AdS$_{d+1}$ at
$r\to\infty$, and $\mu$ is proportional to the Arnowitt-Deser-Misner
(ADM) mass~\cite{atz}.  For $\mu=0$ the
spacetime is locally AdS$_{d+1}$, in unusual coordinates. For
$\mu\ne0$ there is a scalar curvature singularity at $r\to0$.

The global structure and the thermodynamics are analysed 
in \cite{Emparan:1999gf,atz} 
(for certain special cases, see also 
\cite{vanzo-topo,blp,mann-topo,lemos-topo,birmingham}). 
We summarise here the properties that are relevant 
for the rest of the paper, taking care to 
write the formulas for $k\in\mathbb{R}$ 
instead of the conventional normalisation 
to $k=\pm1$ or $k=0$. 

We consider the parameter range in which the spacetime is a
nonextremal black hole,
with the black (and white) hole horizon at
$\hr = \hrp$ and the static exterior region at
$\hrp < \hr < \infty$.
An elementary analysis of the
function
$F$ \nr{eq:Ffunc-def} shows that we can adopt $\hrp$ as the
independent parameter, with the range $\hr_\rmi{ex} < \hrp <
\infty$, where
\be
\hr_\rmi{ex} =
\begin{cases}
0
&
\text{for $k\ge0$},
\\[1ex]
{\displaystyle \sqrt{- \frac{k(d-2)}{d}}}
&
\text{for $k<0$}.
\end{cases}
\label{eq:hrex-def}
\ee
The mass parameter $\mu$ is then given by
\be
\mu = \hrp^d + k \hrp^{d-2},
\label{myy}
\ee
and its range is $\mu_\rmi{ex} < \mu < \infty$, where
\be
\mu_\rmi{ex} =
\hr_\rmi{ex}^d + k \hr_\rmi{ex}^{d-2}
=
\begin{cases}
0
&
\text{for $k\ge0$},
\\[1ex]
{\displaystyle
-{(-k)}^{d/2} \frac{2}{d-2}\left(\frac{d-2}{d}\right)^{d/2} }
&
\text{for $k<0$}.
\end{cases}
\label{eq:muex-def}
\ee

The black hole temperature~$T$,
defined with respect the timelike
Killing vector~$\partial_t$,
is
\be
T= \frac{1}{4\pi \CL}  \left. \frac{dF}{d\hr} \right|_{\hr=\hrp}
=\frac{1}{4\pi\CL}\left(d\,\hrp+\frac{k(d-2)}{\hrp}\right).
\label{T-for-hrp}
\ee
The Bekenstein-Hawking entropy $S$ equals $1/(4 G_{d+1})$ times the
horizon area,
\be
S = \frac{{(\CL \hrp)}^{d-1} \, \Omega_{d-1}(k) }{4G_{d+1}}  .
\label{S-for-hrp}
\ee
Keeping $\Omega_{d-1}(k)$ fixed and allowing $\hrp$ to vary, it is
seen from
\nr{myy}, \nr{T-for-hrp} and
\nr{S-for-hrp} that the first law of black hole thermodynamics takes
the form
\be
\frac{(d-1)\CL^{d-2} \, \Omega_{d-1}(k)}{16\pi G_{d+1}}
\, d\mu
=T dS.
\label{eq:firstlaw-bulk}
\ee

Note that for $k\ge0$ the limit $\hrp \to \hr_\rmi{ex}$ yields a
locally AdS spacetime, but for $k<0$ this limit yields an extremal
black hole with $T=0$ and $S>0$. Note also that the relative sign
of the two terms in \nr{T-for-hrp} changes with the sign
of~$k$. $T$~has a local minimum as a function of $\hrp$ for $k>0$, and
this minimum leads to the Hawking-Page phase
transition~\cite{hp,witten}, but no such phase transition occurs for
$k\le0$~\cite{birmingham}.

\section{Static FRW as the conformal boundary}
\label{sec:static}

In this section we write out the holographic energy-momentum tensor
for the bulk solution \nr{adsbh} when the infinity is approached via
surfaces of constant~$r$.
The results agree with those found in 
\cite{bala-kraus,Emparan:1999gf} in the counterterm formalism of
Balasubramanian and Kraus~\cite{bala-kraus}, but we include here some
of the computational steps as done in the counterterm formalism of de Haro,
Skenderis and Solodukhin~\cite{skenderis}, since comparison with this
computation will allow a concise treatment of the nonstatic foliation
in Section~\ref{sec:tau-in-bulk}. 

To transform the metric \nr{adsbh} to the canonical form~\nr{fefferman},
we must
integrate
$dr/\sqrt{F} = - \CL dz/z$,
or equivalently
\be
\frac{d\hr}{\sqrt{F}} = - \frac{d\hz}{\hz},
\label{eq:fg-static-diffeq}
\ee
where $\hz = z/\CL$ and the boundary condition is
\be
\textrm{$\hz \hr \to 1$ as $\hr\to\infty$} .
\label{eq:BC}
\ee
It follows that the conformal boundary metric is
\be
g^{(0)}_\mn
=
\mathrm{diag}
\left(-1, \CL^2 h_{ij}\right) .
\label{eq:FRW-generald}
\ee
The boundary spacetime is thus the static $d$-dimensional FRW
spacetime,
\be
ds^2_{\mathrm{static}}
= - dt^2 + \CL^2 d \Omega_{d-1}^2(k) ,
\label{eq:staticb}
\ee
with the spatial curvature parameter $k$ and the constant
scale factor~$\CL$.

In the special case $k=0$, equation \nr{eq:fg-static-diffeq}
integrates to
\be
\hr^{d/2}=\frac{1}{\hz^{d/2}}
\left(1 +\frac{\mu}{4} \hz^{d}\right) ,
\ee
and the bulk metric takes the form
\be
ds^2=
\frac{\CL^2}{z^2}
\left\{
-
\frac{\displaystyle \left( 1 - \frac{\mu \hz^d}{4}
\right)^2}
{\displaystyle \left( 1 + \frac{\mu\hz^d}{4}
\right)^{\frac{2(d-2)}{d}}}
\, d\tau^2
+
{\left( 1 + \frac{\mu\hz^d}{4}
\right)^{\frac{4}{d}}}
\CL^2 \, d\Omega^2_{d-1}(0)
+ dz^2
\right\} .
\label{eq:staticfg-k=0}
\ee
We shall consider general values of~$k$. We look first at
arbitrary odd~$d$,
where it suffices to use~\nr{Tmunu-odd}.
The case of even $d$ is more complicated because
of the conformal anomaly~\cite{skenderis},
and we consider only $d=4$
and $d=6$.

\subsection{Odd $d$}

Let $d$ be odd. The energy-momentum tensor
\nr{Tmunu-odd}
is then proportional to the coefficient $g^{(d)}_\mn$
in~\nr{gexp}. To evaluate this coefficient,
we observe from \nr{eq:BC}
that the term in $F$ \nr{eq:Ffunc-def}
proportional to $\mu$ contributes to
$g^{(d)}_{tt}$ by $\mu/\CL^d$.
From \nr{eq:BC} it further follows that all the
other contributions to $g^{(d)}_\mn$
are proportional to $g^{(0)}_\mn$,
and the coefficient of these contributions
is fixed by observing that $T_\mn$
must be traceless.
We hence have
\be
T_\mu{}^\nu
= \frac{\CL^{d-1}}{4 \pi G_{d+1}}
\frac{\mu}{4\CL^d}
\,\,
\mathrm{diag}
\left(
1-d, 1, 1, \ldots, 1
\right) .
\label{Tmn_oddd}
\ee

\subsection{$d=4$}

Let $d=4$. The transformation between $\hr$ and $\hz$
reads
\be
\hr^2=\frac{1}{\hz^2}\left(1 - \frac{k\hz^2}{2}+
\frac{\hz^4}{\hz_+^4}
\right) ,
\ee
where
\be
\hz_+ = \frac{2}{\sqrt{2\hr_+^2 + k}} ,
\label{zhp-def-4}
\ee
and the bulk metric becomes
\be
ds^2=\frac{\CL^2}{z^2}
\left[-
\frac{\displaystyle \left( 1 - \frac{\hz^4}{\hz_+^4}
\right)^2}
{\displaystyle \left( 1 - \frac{k \hz^2}{2}+
\frac{\hz^4}{\hz_+^4}
\right)}
\, dt^2 +
\left( 1 - \frac{k \hz^2}{2}+
\frac{\hz^4}{\hz_+^4}
\right)
\CL^2 \, d\Omega_3^2(k)
+ dz^2
\right] ,
\label{staticmetric-4k}
\ee
Note that $\hz_+$ is always positive.
The exterior region is at $0 < \hz < \hz_+$, the infinity is at
$\hz\to0$ and the horizon is at $\hz\to\hz_+$. If desired, we can
replace $\hr_+$ by $\hz_+$ as the independent parameter:
the range of $\hz_+$ is then
$0 < \hz_+ < 2/\sqrt{k}$ for $k>0$ and $0 < \hz_+ < \infty$ for
$k\le0$. The expression for $\mu$ in terms of $\hz_+$ is
\be
\mu = \frac{4}{\hz_+^4} - \frac{k^2}{4} .
\label{eq:mu-ito-hz+k}
\ee

The formula for the holographic energy-momentum tensor
for $d=4$ reads \cite{skenderis}
\begin{align}
T_\mn=\frac{\CL^3}{4\pi G_5}\left[g^{(4)}_{\mu\nu}
-\fra18 g^{(0)}_\mn[(\Tr g_{(2)})^2-\Tr
   (g_{(2)}^2)]
   - \fra12 (g_{(2)}g_{(0)}^{-1}g_{(2)})_\mn
   + \fra14 (\Tr g_{(2)}) \,
   g_{(2)\mn} \right] .
\label{eq:Tmunu-d=4}
\end{align}
A direct computation yields
\begin{align}
T_\mu{}^\nu&={1 \over4\pi G_5 \CL\hz_+^4}
\,\,
\mathrm{diag}
\left(
-3, 1, 1, 1
\right)
\nonumber
\\
&= \frac{\CL^3}{4\pi G_5} \frac{\mu+\frac14k^2}{4\CL^4}
\,\,
\mathrm{diag}
\left(
-3, 1, 1, 1
\right) .
\label{tmunustatic-4}
\end{align}

\subsection{$d=6$}

Let $d=6$. The expansion \nr{gexp} reads
\ba
g_\mn dx^\mu dx^\nu & = &
- \left[
1 + \fra12 k \hz^2+ \fra{1}{16} k^2 \hz^4 - \fra56 \mu \hz^6 + O(\hz^8)
\right] dt^2
\nonumber
\\
&&
+
\left[
1 - \fra12 k \hz^2 + \fra{1}{16} k^2 \hz^4 + \fra16 \mu \hz^6 + O(\hz^8)
\right]
\CL^2 d \Omega_{5}^2(k) .
\ea
Using the formulas in~\cite{skenderis}, we find
\be
T_\mu{}^\nu=
{\CL^5 \over4\pi G_7}{\mu-\frac18 k^3 \over 4\CL^6}
\,\,
\mathrm{diag}
\left(
-5, 1, 1, 1, 1, 1
\right) .
\label{tmunustatic6}
\ee

\subsection{Comparison} 
\label{subsec:T-d-comparison} 

When the bulk is pure AdS, $\mu=0$, formulas 
\nr{Tmn_oddd}, \nr{tmunustatic-4} and
\nr{tmunustatic6} show that $T_\mn$ 
vanishes in odd dimensions for all $k$ 
but in $d=4$ and $d=6$ only for $k=0$. 

For $k\ne0$, there are three qualitative differences between 
$d=4$ and $d=6$: 
\begin{itemize} 
\item
For $k>0$, the signs of the correction terms in 
\nr{tmunustatic-4} and 
\nr{tmunustatic6} differ. 
When the bulk is pure AdS, $T_{00}$ is positive in $d=4$ 
but negative in $d=6$. 
\item 
For $k<0$, 
the signs of the correction terms in 
\nr{tmunustatic-4} and 
\nr{tmunustatic6} agree, 
so that 
for pure AdS bulk $T_{00}$ is positive both in $d=4$ and in $d=6$. 
However, in the extremal black hole limit, 
$\mu \to \mu_\rmi{ex}$, 
$T_\mn$ vanishes in $d=4$ but remains  
nonvanishing, with positive $T_{00}$, in $d=6$. 
\item 
Let $k$ have either sign. In $d=4$,  
$T_\mn$ is nonvanishing 
everywhere in the (nonextremal, for $k<0$) 
black hole parameter range. 
In $d=6$, 
$T_\mn$ changes sign at 
$\mu = \frac18 k^3$, 
or $\hrp = \frac12 \sqrt{|k|\left(\sqrt{5} - \sgn(k)  \right)}$, 
which is in the black hole parameter range. 
Note that for $k<0$ this value is at negative~$\mu$, 
between pure AdS and the extremal black hole limit. 
We are not aware of 
geometrically special features of the $d=6$ bulk 
at this value of~$\mu$. 
\end{itemize}

\section{Thermodynamics on static FRW}
\label{sec:static-thermo}

In this section we discuss the thermodynamics of the
holographic energy-momentum tensor in the metric
$ds^2_{\mathrm{static}}$~\nr{eq:staticb}.
We take $d$ to be either odd or equal to $4$ or~$6$, with the
energy-momentum tensor given respectively by
\nr{Tmn_oddd}, \nr{tmunustatic-4}
or~\nr{tmunustatic6}.

To begin, we must decide whether
to attribute all of
$T_\mn$ to a fluid or whether $T_\mn$ could contain also a
Casimir-type contribution.
For the reasons discussed in the Introduction, we write 
\be
T_\mn =
T_{\mu\nu}^{(\mathrm{Casimir})}
+
T_{\mu\nu}^{(\mathrm{fluid})} ,
\label{eq:T-split}
\ee
where $T_{\mu\nu}^{(\mathrm{Casimir})}$ is the limiting value of
$T_\mn$ when the bulk black hole vanishes for
$k\ge0$ and becomes extremal for $k<0$,
\begin{subequations}
\ba
\left(T^{(\mathrm{Casimir})}\right)_\mu{}^\nu
&=& \frac{\CL^{d-1}}{4 \pi G_{d+1}}
\frac{\mu_\rmi{ex}}{4\CL^d}
\,\,
\mathrm{diag}
\left(
1-d, 1, 1, \ldots, 1
\right) ,
\qquad \textrm{(for $d$ odd)}
\label{eq:Tmn-static-Casimir-oddd}
\\
\noalign{\vspace{.7ex}}
\left(T^{(\mathrm{Casimir})}\right)_\mu{}^\nu
&=&
\frac{\CL^3}{4\pi G_5} \frac{\mu_\rmi{ex}+\frac14k^2}{4\CL^4}
\,\,
\mathrm{diag}
\left(
-3, 1, 1, 1
\right) ,
\qquad \textrm{(for $d=4$)}
\label{eq:Tmn-static-Casimir-d=4}
\\
\noalign{\vspace{.7ex}}
\left(T^{(\mathrm{Casimir})}\right)_\mu{}^\nu
&=&
{\CL^5 \over4\pi G_7}{\mu_\rmi{ex}-\frac18 k^3 \over 4\CL^6}
\,\,
\mathrm{diag}
\left(
-5, 1, 1, 1, 1, 1
\right) .
\qquad \textrm{(for $d=6$)}
\label{eq:Tmn-static-Casimir-d=6}
\ea
\end{subequations}
The remainder
is given by
\be
\left(T^{(\mathrm{fluid})}\right)_\mu{}^\nu
= \frac{\CL^{d-1}}{4 \pi G_{d+1}}
\frac{(\mu - \mu_\rmi{ex})}{4\CL^d}
\,\,
\mathrm{diag}
\left(
1-d, 1, 1, \ldots, 1
\right) .
\label{eq:Tmn-static-fluid}
\ee
We interpret
$T_{\mu\nu}^{(\mathrm{Casimir})}$
as the vacuum energy-momentum and
$T_{\mu\nu}^{(\mathrm{fluid})}$
as the energy-momentum due to a fluid. By construction,
$T_{\mu\nu}^{(\mathrm{fluid})}$ then vanishes
when the bulk black hole vanishes for
$k\ge0$ and becomes extremal for $k<0$.

$T_{\mu\nu}^{(\mathrm{fluid})}$
is of the perfect fluid form,
\be
T_{\mu\nu}^{(\mathrm{fluid})}
= (\epsilon + p) u_\mu u_\nu + p g^{(0)}_\mn ,
\label{T-perffluid}
\ee
where
$u^\mu = {(\partial_t)}^\mu  = (1,0,\ldots,0)$ and
\be
p = \frac{\CL^{d-1}}{4 \pi G_{d+1}}
\frac{(\mu-\mu_\rmi{ex})}{4\CL^d} ,
\qquad
\epsilon = (d-1) p .
\label{eq:eps-p-static}
\ee
As $\partial_t$
is induced by the bulk Killing vector with respect to
which we defined the Hawking temperature
in Section~\ref{sec:ads-schw-arb},
the bulk black hole induces for the boundary fluid the temperature
$T$ given by \nr{T-for-hrp} and the entropy density $s$ given from
\nr{S-for-hrp} by
\be
s=
\frac{S}{V}
=
\frac{\hrp^{d-1}}{4G_{d+1}} ,
\label{s-for-hrp}
\ee
where
$V = \CL^{d-1} \Omega_{d-1}(k)$ is the spatial volume.

Now, can the boundary fluid
be interpreted as a fluid in local thermal equilibrium?
The fluid certainly satisfies the first law of thermodynamics.
If the spatial volume
$V$ is held constant, $k$~is a constant,
the only independent variable is~$\hrp$,
and equations \nr{eq:firstlaw-bulk},
\nr{eq:eps-p-static} and \nr{s-for-hrp} show that the
first law reads
\be
d \epsilon =T ds.
\ee
If also $V$ is regarded as a thermodynamical variable,
$k$ is determined in terms of $V$ by
$k = \sgn(k) {(V_0/V)}^{2/(d-1)}$,
where
\be
V_0 =
\begin{cases}
\CL^{d-1} \Omega_{d-1}(k/|k|)
&
\text{for $k\ne0$},
\\
0
&
\text{for $k=0$}. 
\end{cases}
\ee
The first law can then be verified to hold in the usual form
\be
dE = T dS - p dV ,
\label{eq:firstlaw-k}
\ee
where $E = V\epsilon$ is the total energy.
The thermodynamical potential $E(S,V)$ in the
microcanonical ensemble can be written in the form
\ba
E(S,V) &=& \frac{(d-1) V}{16 \pi G_{d+1} \CL}
\left[
\left(\frac{4 G_{d+1} S}{V}\right)^{d/(d-1)}
+ \sgn(k)
\left(\frac{V_0}{V}\right)^{2/(d-1)}
\left(\frac{4 G_{d+1} S}{V}\right)^{(d-2)/(d-1)}
\right.
\nonumber
\\
&&
\hspace{15ex}
\left.
+ \Theta(-k) \frac{2}{d-2}\left(\frac{d-2}{d}\right)^{d/2}
\left(\frac{V_0}{V}\right)^{d/(d-1)}
\right] ,
\label{eq:fundrel}
\ea
where $\Theta$ is the Heaviside function. 
Note that as 
$\Theta$ and the signum function
appear in \nr{eq:fundrel} with positive powers of $V_0$, the values of these functions at the origin do not affect~\nr{eq:fundrel}. 

Where the interpretation of the boundary matter as a fluid in local
thermal equilibrium breaks down, however,
is that the Helmholtz free energy density
$f \equiv \epsilon - T s$ of such a fluid should
satisfy $f = -p$~\cite{reichl-book}, but for us
this holds only when $k=0$.
From
\nr{T-for-hrp},
\nr{eq:eps-p-static}
and \nr{s-for-hrp}
we find
\be
f+p
=
\frac{2 k \hrp^{d-2} - d \mu_\rmi{ex}}{16 \pi G_{d+1} \CL} ,
\label{eq:f+p:static}
\ee
and from
\nr{eq:hrex-def}
and
\nr{eq:muex-def}
it follows that the right-hand side vanishes when $k=0$ but cannot
vanish for any value of $\hrp$ when $k\ne0$.
The physical reason is that the
spatial curvature breaks scale invariance:
a simultaenous scaling of $E$, $S$ and~$V$,
which three
would conventionally be considered the extensive variables,
leaves the fundamental thermodynamic relation
\nr{eq:fundrel} invariant only for $k=0$. For $k\ne0$,
scale invariance holds asymptotically in
the large volume limit,
in which the second and third term in
\nr{eq:fundrel} are small compared with the first term.
From the bulk point of view this is the limit of
a black hole with large ADM mass.

We emphasise that this thermodynamic breakdown for
$k\ne0$ is not in the first law of
thermodynamics but in the phenomenological description of the boundary
matter as a thermalised fluid. That such a breakdown can be expected
when the characteristic thermal wave length $1/T$ is not small
compared with the curvature length scales was noted already
in~\cite{Gibbons-Hawking}, and in the AdS/CFT context it has been
observed for example in~\cite{ss}.

We end the section with four comments.
First, if $V$ is held constant, $k$ is a constant,
and the first law for the
Helmholtz free energy density reads $df = - s \,
dT$. As a consistency check on the thermodynamics, we
should therefore
be able to obtain $f$ also by evaluating the partition function of
the canonical ensemble. In AdS/CFT
duality, the semiclassical approximation to this partition function is
$Z_\rmi{CFT}=e^{-I}$, where $I$ is the minimal supergravity action
appropriate for fixing the induced boundary metric, in positive
definite signature, and the temperature is identified as the inverse
period of the imaginary time in the boundary metric.
For definiteness, we specialise to $d=4$.
The regularised Lorentz-signature Einstein action 
is given in Appendix B of \cite{skenderis} and reads 
\be
S={1\over16\pi G_5}\left\{\int d^5x\sqrt{-g}\,\left({8\over\CL^2}\right)
-\int d^4 x\sqrt{-\gamma}
\left[2K - {6\over\CL}+{\CL\over 2}R(\gamma)
+\textrm{(log term)}
\right]_{\hz=\heps}\right\},
\label{regaction}
\ee
where $\gamma_\mn$ is the induced metric on the surface $\hz=\heps$,
$K$ is the extrinsic curvature of this surface, the logarithmic term
is given in~\cite{skenderis}, and
(unlike elsewhere in the present paper)
all the sign conventions are those
of~\cite{skenderis}. Inserting the metric \nr{staticmetric-4k}
in~\nr{regaction}, the integration over the coordinates $y^i$ gives
the factor $\CL^3\Omega_3(k) = V$, and the
integration over $t$ gives, when continued to positive definite
signature, the factor~$1/T$.  The remaining integration in the volume
integral part is
\be
\int_{\hz = \heps}^{\hz = \hz_+}{d\hz\over\hz^5}
\left(1-{\hz^4\over\hz^4_+}\right)
\left(1 - {k \hz^2\over2}+{\hz^4\over\hz^4_+}\right)
={1\over4\heps^4}-{k\over4\heps^2} +
{k\hz_+^2 -1 \over 2\hz_+^4}+ \mathcal{O}(\heps^2) .
\label{eq:regaction-eval1}
\ee
The boundary term in
\nr{regaction} can be verified to cancel the divergent terms in
\nr{eq:regaction-eval1}
and to bring in no contributions that remain finite as
$\heps\to0$. From the finite
term in \nr{eq:regaction-eval1} we then find, after continuation to positive
definite signature, that
\be
I = \frac{1}{4\pi G_5 \CL}
\left( \frac{V}{T}\right)
\frac{k\hz_+^2 -1}{\hz_+^4}
= \bigl(V/T\bigr) f_{\mathrm{tot}} ,
\label{eq:I-regularised}
\ee
where $f_{\mathrm{tot}} \equiv \epsilon_{\mathrm{tot}} - T s$
and the total energy density
$\epsilon_{\mathrm{tot}} \equiv
T_{00} =
T_{00}^{(\mathrm{Casimir})}
+
T_{00}^{(\mathrm{fluid})}$
includes both the fluid part and the Casimir part.
Note that
$f_{\mathrm{tot}}$ and $f$ differ only for $k>0$.
This is the consistency check we sought for.

We note in passing that the action computed in \cite{Burgess:1999vb}
by a volume subtraction method agrees with our action
\nr{eq:I-regularised} only for $k=0$.  For $k\ne0$, the difference
affects the large temperature expansion of the Helmholtz free
energy so that the coefficient of the $k^2$ term in equation (43) in
\cite{Burgess:1999vb} is halved.

Second, one may ask to what extent our conclusions depend on 
splitting the energy-momentum tensor into the 
Casimir part and the fluid part as in \nr{eq:T-split}. 
It can be verified that the consistency of the first law for any $k$
and the breaking of
scale invariance for $k\ne0$ 
hold 
for $d=4$ and $d=6$
even if we choose the Casimir term so that the 
fluid energy-momentum tensor vanishes for $\mu=0$, 
and they hold for any $d$ even if we choose the Casimir term to vanish. 

Third, for $k<0$ the split \nr{eq:T-split} implies that
the energy density, the pressure and the
temperature all go to zero in the
extremal black hole limit,
but the entropy density does not.  In a
fluid interpretation, this would mean that the fluid becomes highly
degenerate in the zero temperature limit. It has however been argued
that in this limit thermal fluctuations become so large that a
statistical description of the system is no longer
viable~\cite{Preskill:1991tb}.

Fourth, we note that for $d=4$, the fundamental relation
\nr{eq:fundrel}
with $k=0$
has the same structure as the leading term in the
fundamental relation obtained in \cite{ss} for boundary hydrodynamics
in the presence R-charges in the bulk.

\section{Linearly expanding FRW as the conformal boundary}
\label{sec:tau-in-bulk}

In this section we write the generalised Schwarzschild-AdS$_{d+1}$
black hole \nr{adsbh}
in coordinates in which the conformal
boundary is a linearly expanding FRW cosmology.

Consider the metric~\nr{adsbh-justmetric}, and allow to begin with
$F$ to be an arbitrary positive function of $r$ only.
We replace $(t,r)$ 
by the positive-valued coordinates 
$(\tau,z)$ of dimension length by 
\begin{subequations}
\label{eq:rttransf}
\ba
\frac{dr}{\CL\sqrt{F+1}} 
&=& 
- d \left[ 
\log \left(\frac{z}{\tau}\right) 
\right] , 
\label{eq:rtransf}
\\[1ex]
\frac{dt}{\CL} 
&=& 
\frac{d\tau}{\tau} 
- \frac{1}{F} \, 
d \left[ 
\log \left(\frac{z}{\tau}\right) 
\right] . 
\ea
\end{subequations} 
The metric becomes 
\be
ds^2
=
\frac{\CL^2}{z^2}
\left[ - (F+1) \left(\frac{z}{\tau}\right)^2  d\tau^2
+ \left(\frac{rz}{\CL\tau}\right)^2
\tau^2 d\Omega_{d-1}^2(k)
+ dz^2
\right] ,
\label{eq:tau11metric}
\ee
where it follows from \nr{eq:rtransf}
that $r$ and $F$ depend on the coordinates $(\tau,z)$ only
through the combination~$z/\tau$. The Killing vector
$\partial_t$ of \nr{adsbh-justmetric} takes in \nr{eq:tau11metric} the
form $\CL^{-1}(\tau\partial_\tau + z\partial_z)$. Note that $z$
differs from the coordinate $z$ used in
Section~\ref{sec:static}.

To proceed, suppose that 
$F = {(r/\CL)}^2 + O(1)$ at $r\to\infty$, as the case
is in~\nr{eq:Ffunc-def}, and choose the integration constant in
\nr{eq:rtransf} so that 
$z/\tau = (\CL/r) + O\left[ (\CL/r)^3\right]$ at
$r\to\infty$. The metric \nr{eq:tau11metric} has then at $z\to0$ the
Fefferman-Graham form of \nr{fefferman} and~\nr{gexp}, with
the boundary metric
\be
g^{(0)}_\mn
=
\mathrm{diag}
\left(-1, \tau^2 h_{ij}\right) .
\label{eq:FRW-expanding}
\ee
The boundary spacetime is thus the $d$-dimensional FRW cosmology with
the spatial curvature parameter $k$ and the scale factor~$\tau$,
\be
ds^2_{\mathrm{exp}}
= - d\tau^2 + \tau^2 d \Omega_{d-1}^2(k) .
\label{eq:expandingb}
\ee
As reviewed in the Appendix, $ds^2_{\mathrm{exp}}$
has a scalar curvature singularity at $\tau\to0$ for $k\ne-1$,
but for $k=-1$ it is flat and its comoving
world lines are those of an isotropic and
boost-invariant explosion in Minkowski space.

We now specialise to $F$ given by~\nr{eq:Ffunc-def}.
Comparing \nr{eq:rtransf}
and
\nr{eq:tau11metric} with respectively 
\nr{eq:fg-static-diffeq}
and 
\nr{adsbh}
shows that the 
metric \nr{eq:tau11metric} is obtained from the 
Fefferman-Graham bulk metric 
of Section \ref{sec:static} by keeping the overall factor
$\CL^2/z^2$ and elsewhere just making the replacements $t \to \tau$, $k\to
k+1$ and $\CL \to \tau$. When $d$ is arbitrary and
$k=-1$, we in particular have from
\nr{eq:staticfg-k=0}
\be
ds^2=\frac{\CL^2}{z^2}
\left\{-
\frac{\displaystyle \left[ 1 - \frac{\mu}{4}\left(\frac{z}{\tau}\right)^d
\right]^2}
{\displaystyle \left[ 1 + \frac{\mu}{4}\left(\frac{z}{\tau}\right)^d
\right]^{\frac{2(d-2)}{d}}}
\, d\tau^2+
{\left[ 1 + \frac{\mu}{4}\left(\frac{z}{\tau}\right)^d
\right]^{\frac{4}{d}}}
\tau^2 \, d\Omega_{d-1}^2(-1)+ dz^2
\right\} ,
\label{eq:tau-metric}
\ee
and when $d=4$ and $k$ is arbitrary,
we have from \nr{staticmetric-4k}
\ba
ds^2 &=&
\frac{\CL^2}{z^2}
\left\{-
\frac{\displaystyle
\left[ 1 - \left( \frac{\mu + \frac14 {(k+1)}^2}{4} \right)
\frac{z^4}{\tau^4}
\right]^2}
{\displaystyle \left[ 1 - \frac{(k+1)z^2}{2\tau^2}+
\left(\frac{\mu + \frac14 {(k+1)}^2}{4} \right) \frac{z^4}{\tau^4}
\right]}
\, d\tau^2
\right.
\nonumber
\\
&&
\hspace{5ex}
\left.
\vphantom{
\frac{\displaystyle
\left[ 1 - \left(\frac{\mu + \frac14 {(k+1)}^2}{4}
\right) \frac{z^4}{\tau^4}
\right]^2}
{\displaystyle \left[ 1 - \frac{(k+1)z^2}{2\tau^2}+
\left(\frac{\mu + \frac14 {(k+1)}^2}{4} \right) \frac{z^4}{\tau^4}
\right]}
}
+
\left[ 1 - \frac{(k+1)z^2}{2\tau^2}+
\left(\frac{\mu + \frac14 {(k+1)}^2}{4} \right) \frac{z^4}{\tau^4}
\right]
\tau^2 \, d\Omega_3^2(k)
+ dz^2
\right\} .
\label{expmetric-4k}
\ea
For $k=0$, the metric \nr{expmetric-4k} reduces to the case
$r(\tau)=\tau/\tau_0$ of the bulk solution (17)--(18) of~\cite{kt}.
The loci of the coordinate singularities in these nonstatic
forms of the bulk metric can be analysed
as in~\cite{klt},
for example by adapting the
Eddington-Finkelstein techniques of~\cite{Kinoshita:2008dq}, but this
will not be needed in what follows.

The holographic energy-momentum tensor can be computed as in
Section~\ref{sec:static}. We find
\begin{subequations}
\label{eq:Tmn-exp-all}
\ba
T_\mu{}^\nu
&=& \frac{\CL^{d-1}}{4 \pi G_{d+1}}
\frac{\mu}{4\tau^d}
\,\,
\mathrm{diag}
\left(
1-d, 1, 1, \ldots, 1
\right) ,
\qquad \textrm{(for $d$ odd)}
\label{eq:Tmn-exp-odd}
\\
T_\mu{}^\nu
&=& \frac{\CL^3}{4\pi G_5} \frac{\mu+\frac14 {(k+1)}^2}{4\tau^4}
\,\,
\mathrm{diag}
\left(
-3, 1, 1, 1
\right) ,
\qquad \textrm{(for $d=4$)}
\label{tmunuexp-4}
\\
T_\mu{}^\nu
&=&
{\CL^5 \over4\pi G_7}{\mu-\frac18 {(k+1)}^3 \over 4\tau^6}
\,\,
\mathrm{diag}
\left(
-5, 1, 1, 1, 1, 1
\right) .
\qquad \textrm{(for $d=6$)}
\label{tmunuexp6}
\ea
\end{subequations}

As a consistency check, recall that
since both the metric
\nr{eq:tau11metric} and the static metric used in Section
\ref{sec:static} have the asymptotic form of \nr{fefferman}
and~\nr{gexp}, the conformal boundary metrics $ds^2_{\mathrm{static}}$
\nr{eq:staticb} and $ds^2_{\mathrm{exp}}$ \nr{eq:expandingb}
must be related by a conformal transformation. From 
\nr{eq:rttransf}
it follows that on a surface of constant $r$
we have $\tau = ({\mathrm{const}}) \times \exp(t/\CL)$, and the
conformal transformation therefore reads
\be
ds^2_{\mathrm{exp}} =
(\tau^2/\CL^2) \, ds^2_{\mathrm{static}} .
\label{eq:conf-transf}
\ee
For odd~$d$, the energy-momentum tensor in
$ds^2_{\mathrm{exp}}$ should therefore be ${(\CL/\tau)}^d$ times
that in $ds^2_{\mathrm{static}}$, and formulas
\nr{Tmn_oddd}
and
\nr{eq:Tmn-exp-odd}
show that this is indeed the case.
For even $d$ the situation is more complicated since the
transformation includes also the conformal anomaly. For definiteness,
consider $d=4$ with $k=-1$. In this case
the anomaly term that must be added to
\nr{tmunustatic-4} before scaling by ${(\CL/\tau)}^4$
is given in formula (6.139) of \cite{bd} and is
a linear combination of the tensors
\begin{subequations}
\label{threeH}
\ba
{}^{(1)}H_\mu^{\,\,\,\nu}\ &\equiv&
2 R_{;\mu}{}^{;\nu}
- 2 g_\mu{}^{\nu} \Box R
- \fra{1}{2}R^2g_\mu{}^{\nu}
+ 2 R R_{\mu}{}^{\nu}
={6\over\CL^4} \, {\rm diag}(-3,1,1,1),
\\
{}^{(3)}H_\mu^{\,\,\,\nu}\ &\equiv&
\fra{1}{12}R^2g_\mu{}^{\nu}-R^{\alpha}{}_{\beta}
R_{\alpha\mu}{}^{\beta\nu}
={1\over\CL^4} \, {\rm diag}(3,-1,-1,-1),
\ea
\end{subequations}
evaluated for $ds^2_{\mathrm{static}}$ as shown. The anomaly
contribution can thus be made to precisely cancel the term $\frac14$
from $\mu + \frac14$ in~\nr{tmunustatic-4}, and give agreement
with~\nr{tmunuexp-4}, provided the
coefficients of the tensors
\nr{threeH} satisfy an appropriate linear relation.
This linear relation is in particular satisfied by the
anomaly coefficients that occur in
$\mathcal{N}=4$ SU(N) Super-Yang-Mills
theory~\cite{Balasubramanian:2002am}.

\section{Thermodynamics on linearly
expanding FRW: the isotropic fireball}
\label{sec:exp-thermo}

We are now ready to address the thermodynamics of the comoving fluid
in the expanding FRW cosmology \nr{eq:expandingb}.
We take $d$ to be either odd or equal
to $4$ or~$6$. The treatment follows closely that of the case $d=2$
in~\cite{klt}.

As in Section~\ref{sec:static-thermo}, we split
the energy-momentum tensor \nr{eq:Tmn-exp-all}
into a fluid part
$T_{\mu\nu}^{(\mathrm{fluid})}$ and a
Casimir part $T_{\mu\nu}^{(\mathrm{Casimir})}$, so that
$T_{\mu\nu}^{(\mathrm{fluid})}$ vanishes
when the bulk black hole vanishes for
$k\ge0$ and becomes extremal for $k<0$.
The result is
\be
\left(T^{(\mathrm{fluid})}\right)_\mu{}^\nu
= \frac{\CL^{d-1}}{4 \pi G_{d+1}}
\frac{(\mu - \mu_\rmi{ex})}{4\tau^d}
\,\,
\mathrm{diag}
\left(
1-d, 1, 1, \ldots, 1
\right)
\label{eq:Tmn-exp-fluid}
\ee
and
\begin{subequations}
\label{eq:Tmn-exp-all-vac}
\ba
\left(T^{(\mathrm{Casimir})}\right)_\mu{}^\nu
&=& \frac{\CL^{d-1}}{4 \pi G_{d+1}}
\frac{\mu_{\rmi{ex}}}{4\tau^d}
\,\,
\mathrm{diag}
\left(
1-d, 1, 1, \ldots, 1
\right) ,
\qquad \textrm{(for $d$ odd)}
\label{eq:Tmn-exp-odd-vac}
\\
\left(T^{(\mathrm{Casimir})}\right)_\mu{}^\nu
&=& \frac{\CL^3}{4\pi G_5} \frac{\mu_{\rmi{ex}}+\frac14 {(k+1)}^2}{4\tau^4}
\,\,
\mathrm{diag}
\left(
-3, 1, 1, 1
\right) ,
\qquad \textrm{(for $d=4$)}
\label{tmunuexp-4-vac}
\\
\left(T^{(\mathrm{Casimir})}\right)_\mu{}^\nu
&=&
{\CL^5 \over4\pi G_7}{\mu_{\rmi{ex}}-\frac18 {(k+1)}^3 \over 4\tau^6}
\,\,
\mathrm{diag}
\left(
-5, 1, 1, 1, 1, 1
\right) .
\qquad \textrm{(for $d=6$)}
\label{tmunuexp6-vac}
\ea
\end{subequations}
As a consistency check on the split, we observe that for $d=4$ with
$k=-1$, our $T_{\mu\nu}^{(\mathrm{Casimir})}$ \nr{tmunuexp-4-vac} does
have the form of the energy-momentum tensor of a conformal scalar
field in the vacuum whose positive frequency modes are conformally
related to those that define the static vacuum on the static universe
\nr{eq:staticb} \cite{bunch-hyperbolic}.

$T_{\mu\nu}^{(\mathrm{fluid})}$ \nr{eq:Tmn-exp-fluid} has the perfect
fluid form \nr{T-perffluid} with a comoving fluid, $u^\mu =
(1,0,\ldots,0)$,
and
\be
p = \frac{\CL^{d-1}}{4 \pi G_{d+1}}
\frac{(\mu-\mu_\rmi{ex})}{4\tau^d} ,
\qquad
\epsilon = (d-1) p .
\label{eq:eps-p-exp}
\ee
The energy density and pressure \nr{eq:eps-p-exp} hence come from
those in the static case \nr{eq:eps-p-static} by scaling with the
factor~$\CL^d/\tau^d$, for both odd and even~$d$: this follows
from the conformal transformation properties of the stress-energy
tensor and from our having grouped the conformal anomalies for even
$d$ in $T_{\mu\nu}^{(\mathrm{Casimir})}$. To define other
time-dependent thermodynamical quantities, we similarly scale those of
the static case by the dimensionally appropriate power of~$\CL/\tau$,
so that $s$ is scaled by $\CL^{d-1}/\tau^{d-1}$ and $T$
by~$\CL/\tau$. This gives
\be
T=
\frac{1}{4\pi\tau}\left(d\,\hrp+\frac{k(d-2)}{\hrp}\right) ,
\qquad s
= \frac{\CL^{d-1}}{4G_{d+1}}\frac{\hr_+^{d-1}}{\tau^{d-1}} .
\label{T-s-exp}
\ee

It is readily verified that the expanding
fluid satisfies the first law of
thermodynamics, in two distinct senses. The volume on the spatial
hypersurface of constant $\tau$ equals $V = \tau^{d-1}
\Omega_{d-1}(k)$, and we can define the total energy and entropy at
given $\tau$ in terms of the densities $\epsilon$ and $s$, given in
\nr{eq:eps-p-exp} and~\nr{T-s-exp}, by respectively $E = V \epsilon$
and $S= V s$. Now, first, if
$\tau$ is regarded fixed and $\hrp$ and $k$ are
allowed to vary, the computations in Section \ref{sec:static-thermo} go
through without change and show that the first law at fixed $\tau$
holds in the standard form~\nr{eq:firstlaw-k}.
Second,
and perhaps more relevantly for the expanding fluid,
the first law
holds in the form \nr{eq:firstlaw-k} also
when $\hrp$ and $k$ are regarded as
fixed and we follow the expansion of the fluid in~$\tau$.
Note that this second sense of the first law follows just by virtue of the
inverse powers of $\tau$ in $s$ and
$\epsilon$ and the tracelessness
of $T_{\mu\nu}^{(\mathrm{fluid})}$,
and it does not hinge on the details of how the
thermodynamical variables depend on $\hrp$ and~$k$.

As in the static case of Section~\ref{sec:static-thermo},
the interpretation of the boundary matter as a fluid 
in local thermal equilibrium breaks down for $k\ne0$
in that the Helmholtz free energy
density $f \equiv \epsilon - T s$ is not equal to $-p$,
except asymptotically in the limit of large~$\hrp$: 
the value of $f+p$ in the expanding case is given 
by just multiplying the static value \nr{eq:f+p:static}
by~$\CL^d/\tau^d$. 
The physical reason is again that for $k\ne0$ the
nonvanishing spatial curvature breaks the scale invariance in
$E$, $V$ and~$S$.
When the scale invariance does hold, however,
the boundary matter is an 
expanding perfect fluid in local thermal equilibrium.

\section{Conclusions}
\label{sec:conclusions}

We have shown that the AdS/CFT dual of
a linearly expanding $d$-dimensional FRW cosmology,
with $d\ge3$ and an arbitrary spatial curvature parameter~$k$,
is a nonstatic foliation of the
generalised Schwarzschild-AdS${}_{d+1}$
black hole with a horizon of constant curvature~$k$. 
The boundary matter is a perfect fluid that satisfies 
the first law of thermodynamics, 
but it admits a description as a scale-invariant
fluid in local thermal equilibrium only when the 
inverse Hawking temperature is negligible compared 
with the spatial curvature length scale. 

Geometrically, the absence of scale invariance 
for generic values of the parameters
is a direct consequence of the nonvanishing
spatial curvature on the boundary
and should hence perhaps not be surprising. 
However, the absence of the scale invariance
implies that the Helmholtz free energy density 
of the fluid is not equal to minus the pressure,
which means that the fluid does not admit a conventional 
interpretation as a fluid in local thermal equilibrium.
It may be surprising that this absence of a conventional thermalised
fluid interpretation occurs even in the special case of the Milne universe,
where the fluid is just an isotropically and boost-invariantly 
expanding fireball in Minkowski space.
As a side remark, we noted that a similar absence
scale invariance occurs already in the more familiar case of a
static bulk foliation, where the boundary is the static FRW cosmology.

The case $d\ge3$ analysed in the present paper
has both similarities to and differences from
the case $d=2$ analysed in~\cite{klt}.
For $d=2$ the spatial curvature necessarily vanishes, and the
adiabatic expansion found in \cite{klt}
for $d=2$ can be regarded as a direct dimensional
continuation of the adiabatic expansion found
in the present paper
for $d\ge3$ with $k=0$.
However, a difference is that for $d=2$ the
linearly expanding cosmology
does describe a fluid expanding isotropically
in $(1+1)$-dimensional Minkowski spacetime,
whereas for $d\ge3$ the Minkowski space fluid
description requires $k=-1$. The adiabatically
expanding isotropic fireball in $d=2$ Minkowski
space does therefore not generalise into an
adiabatically expanding isotropic fireball in
higher-dimensional Minkowski spaces.

While the spatial isotropy has made our model exactly solvable, the
model contains significantly less structure than the model with boost
invariance in one spatial dimension and translational invariance in
the others~\cite{Heller:2008mb,Kinoshita:2008dq}.  In particular, the
phenomenological interest of our model as a description of expanding
gauge theory plasma, especially in the special case of a fluid
expanding isotropically in Minkowski space, is limited by the absence
of viscosity.  One might attempt to bring in shear viscosity by giving
the bulk black hole angular momentum, but the rotating $d=2$ case
analysed in \cite{klt} (and in which case there is no shear viscosity)
raises doubt as to whether the resulting flow configuration would be
phenomenologically relevant.  A more likely prospect for introducing
viscosity in the isotropic setting might be to include metric
deformations or further bulk fields
\cite{parnachev,benincasa-bs,benincasa-b,kty,kiritsis,gubsernellore}.

\vspace{1cm}
{\it Acknowledgements}.
We thank Veronika Hubeny, Esko Keski-Vakkuri,
Don Marolf, Mukund Rangamani,
Dirk Rischke, Paul Romatschke and Andrei Starinets
for discussions and correspondence and 
Elias Kiritsis, 
Makoto Natsuume
and George Siopsis 
for bringing references to our attention. 
This research has been supported in part by Academy of Finland,
contract number 109720, 
by the National Science Foundation under Grant No.\ HY05-51164, 
by STFC (UK) grant PP/D507358/1
and by NSERC, Canada. 
KK thanks the Institute for Nuclear Theory at the
University of Washington for its hospitality and
the Department of Energy for partial support
during the completion of this work.
JL thanks Helsinki Institute of Physics for
hospitality in the final stages of this work.

\begin{appendix}

\section{Appendix: Spaces of constant curvature
and linearly expanding FRW cosmologies}

In this appendix we collect relevant properties of spaces of constant
curvature and linearly expanding FRW cosmologies. As in the main text,
we assume $d\ge 3$.

Let $d\Omega^2_{d-1}(k)$ denote for $k>0$ the metric on the
$(d-1)$-dimensional sphere of curvature radius $1/\sqrt{k}$, for $k=0$
the Euclidean metric on~$\mathbb{R}^{d-1}$, and for $k<0$ the metric
on the $(d-1)$-dimensional hyperbolic space of curvature radius
$1/\sqrt{-k}$. We denote the coordinates by
$y^i$, $i=1,\ldots,d-1$, and write
the metric
as $d\Omega_{d-1}^2(k) = h_{ij} dy^i dy^j$. $d\Omega_{d-1}^2(k)$ is a
metric of constant curvature~\cite{wolf}, with the Riemann tensor
\be
{}^{(d-1)}R_{ijmn} = k
\left(
h_{im} h_{jn} - h_{in} h_{jm}
\right)  ,
\ee
and the Ricci scalar equals ${}^{(d-1)}R = k (d-1)(d-2)$. The explicit
embeddings in $d$-dimensional
Euclidean space for $k>0$ and $d$-dimensional Minkowski space for
$k<0$ are given in~\cite{weinberg}.
In terms of the cosmological area-radius
coordinate~$r$,
the metric takes the form \cite{weinberg}
\be
d\Omega_{d-1}^2(k)
=
\frac{dr^2}{1 - k r^2} + r^2 d\Omega_{d-2}^2(1) .
\label{eq:area-radius}
\ee

In the main text we encounter the $d$-dimensional linearly expanding
FRW cosmology~\nr{eq:expandingb},
\be
ds^2_{\mathrm{exp}} = - d\tau^2 + \tau^2 d
\Omega_{d-1}^2(k) ,
\label{eq:expandingb-app}
\ee
where $0<\tau<\infty$.
The only nonvanishing components of the Riemann tensor of
$ds^2_{\mathrm{exp}}$
are the fully spatial components, given by
\be
{}^{(d)}R_{ijmn} = \tau^2 (k+1)
\left(
h_{im} h_{jn} - h_{in} h_{jm}
\right),
\ee
and the Ricci scalar equals
${}^{(d)}R = (k+1)(d-1)(d-2)/\tau^2$.
$ds^2_{\mathrm{exp}}$ has hence
a scalar curvature singularity at
$\tau\to0$ when $k\ne-1$ but is
flat when $k=-1$.

In the flat special case $k=-1$, $ds^2_{\mathrm{exp}}$ is known as
(the $d$-dimensional
generalisation of) the Milne universe~\cite{rindler-relbook},
\be
ds^2_{\mathrm{Milne}}
= - d\tau^2 + \tau^2 d \Omega_{d-1}^2(-1) .
\label{eq:Milne}
\ee
$ds^2_{\mathrm{Milne}}$ covers the interior of the future light cone
of the origin in $d$-dimensional Minkowski space, the comoving world
lines of constant $y^i$ are inertial world lines that start from the
origin, and the surfaces of constant $\tau$ are spacelike hyperboloids
of constant proper time $\tau$ from the origin.
$ds^2_{\mathrm{Milne}}$ is thus
adapted to a spherically symmetric and
boost-invariant explosion in Minkowski space, starting at one point.
The transformation between \nr{eq:Milne} and
the usual Minkowski coordinates
can be found in \cite{weinberg,rindler-relbook}.

\end{appendix}

\edoc